\documentclass[runningheads]{llncs}
\usepackage[table,xcdraw]{xcolor} % Consolidated xcolor options
\usepackage{tabularx} % For tables that automatically adjust to the width of the page
\usepackage{caption} % For customizing captions
\usepackage[most]{tcolorbox} % Consolidated tcolorbox options
\usepackage{array} % For defining new column types
\usepackage{fontawesome}
\usepackage{graphicx} % for including graphics
\usepackage{cite} % for bibliography
\usepackage{comment}
\usepackage{longtable}

\begin{document}
\title{Enhancing Productivity with AI During the Development of an ISMS: Case Kempower}
\titlerunning{Enhancing Productivity with AI During the Development of an ISMS}
% Optionally add thanks or funding information within the \thanks{...} in the title
\author{Atro Niemeläinen\inst{1}\orcidID{0009-0001-7208-0869} \and
Muhammad Waseem\inst{2}\orcidID{0000-0001-7488-2577} \and
Tommi Mikkonen\inst{2} \orcidID{0000-0002-8540-9918}}
\authorrunning{A. Niemeläinen et al.}
\institute{Kempower, Lahti, Finland\\
\email{atro.niemelainen@kempower.fi} \and
University of Jyväskylä, Jyväskylä, Finland\\
\email{muhammad.m.waseem@jyu.fi, tommi.j.mikkonen@jyu.fi}}

\maketitle

\begin{abstract}

Investing in an Information Security Management System (ISMS) enhances organizational competitiveness and protects information assets. However, introducing an ISMS consumes significant resources; for instance, implementing an ISMS according to the ISO27001 standard involves documenting 116 different controls. This paper discusses how Kempower, a Finnish company, has effectively used generative AI to create and implement an ISMS, significantly reducing the resources required. This research studies how the use of generative AI can enhance the process of creating an ISMS. We conducted seven semi-structured interviews held with various stakeholders of the ISMS project, who had varying levels experience in cyber security and AI.

\end{abstract}

\section{Introduction}
\label{Introduction}

Many companies are investing in Information Security Management Systems (ISMS) to improve their information security management process. An ISMS refers to a systematic approach for establishing, implementing, operating, monitoring, reviewing, maintaining, and improving an organization's information security to achieve business objectives \cite{iso27001}.This also enhances the organization's competitiveness and protects their information assets \cite{Park2010}. Implementing an ISMS not only helps organizations to manage, reduce and identify any information security threats but it also supports the organization's business continuity \cite{Dedy2018}. Moreover, the implementation of an ISMS enhances trust among business partners, significantly boosting business operations, and it creates a more secure working environment by reducing the number of security breaches significantly \cite{jalil2003isms}. 

Existing studies have explored various aspects of integrating AI and multi-agent systems into software development and related fields. For instance, enhancing bid evaluation through an AI procurement assistant has been investigated by \cite{waseem2023artificial}. The role of ChatGPT as a software development bot was studied by \cite{waseem2024chatgpt}, while \cite{rasheed2023autonomous} presented a vision paper on autonomous agents in software development. Additionally, \cite{rasheed2024can} examined the potential of large language models as data analysts, and introduced Codepori for autonomous software development \cite{rasheed2024codepori}. Enabling liquid artificial intelligence with IDSA and GAIA-X was discussed by \cite{waseem2024enabling}. Sami et al. \cite{sami2024system} proposed a system for systematic literature reviews using multiple AI agents. Evaluation of large language models via multi-AI agents was presented by \cite{rasheed2024large}. The Holon programming model for adaptive systems and its enhancement with natural language processing was explored by \cite{ashfaq2024holon} and \cite{ashfaq2024enhancing}. Prioritizing software requirements using large language models was demonstrated by \cite{sami2024prioritizing}. AI-powered code reviews were investigated by \cite{rasheed2024ai}, while experimenting with multi-agent software development platforms was discussed by \cite{sami2024experimenting}. Finally, \cite{malik2024tool} developed a tool for generating test case scenarios using large language models. 
However, introducing an ISMS in an organization consumes a lot of resources, and to the best of our knowledge, we have not found any study that explores the increase in productivity with AI during the development of an ISMS. These include time and money, but also hidden costs as employees become  “tied up" in a lengthy project, leaving them little time for other tasks. Implementing an ISMS according to the ISO27001 standard \cite{iso27001} involves documenting 116 different controls. It is hard to imagine that someone could write all the 116 different control documents without searching for information or using external consultants help. Utilizing generative AI in the creation of an ISMS significantly reduces many of the mentioned issues. It can generate the first draft for each of the 116 documents, thereby eliminating the need for information search during the document creation process. External consultants are not needed since generative AI can create templates for ISO27001 compliant documents. However, a professional must always review the generated document and modify the information to match the company's work practices and organizational structure.

In this paper, we study how Kempower, a Finnish company, has used generative AI (ChatGPT) to help with creating an ISMS, significantly reducing the necessary resources. Our goal is to explore how leveraging AI can enhance ISMS creation within an organization. Practically, this involves constructing ISMS documentation at Kempower using a novel AI-driven method.

This research has three major contributions which are:

\begin{itemize}
    \item Introducing a novel AI-driven approach for ISMS documentation that reduces the resources required for implementation and demonstrates the practical application of generative AI in security management.
    \item Providing insights into how AI can enhance administrative efficiency, particularly in the creation of compliance documents, improving both process speed and document accuracy.
    \item  Illustrating through the Kempower case study how AI integration can increase employee awareness and engagement with ISMS, promoting a technology-forward organizational culture.
\end{itemize}

The rest of this paper is structured as follows: Section 2 discusses the background of this study. Section 3 outlines the case study as the research methodology. Section 4 reports the study results. Section 5 discusses the key findings regarding Kempower's ISMS project and AI usage. Section 6 concludes the study and suggests directions for future work.

\section{Background}
\label{Background}
An ISMS encompasses policies and procedures for systematically managing an organization's sensitive data. By implementing an ISMS, organizations can minimize risks and ensure business continuity by proactively limiting the impact of security breaches. This not only enhances organizational competitiveness and protects information assets \cite{Park2010}, but it also supports the overall business continuity \cite{Dedy2018}.

Organizations that have implemented an ISMS experience several benefits, as summarized in \cite{Park2010}. These include a rise in sales and overall worth due to ISMS certification, cost savings that positively impact financial appreciation, and new business opportunities that enhance transaction stability. Additionally, improved operational efficiency within ISMS-certified enterprises increases trust among stakeholders, while the system’s protective capacity boosts security awareness across all operations.

As with any information system, it requires reliable, timely operations. To that end, artificial intelligence is offering novel opportunities. These include an increased level of automation, as AI can take over complicated, tiresome tasks from human users, as well as improve risk management in the process. However, each organization needs to find their own processes and practices to best benefit from AI and its automation capabilities.

%Benefits for organizations that have implemented an ISMS include a number of benefits, as summarized in \cite{Park2010}. For instance, the rise in sales for an enterprise with an ISMS certification will lead to an increase in its overall worth, and cost savings achieved by an enterprise that holds ISMS certification will have a positive impact on its overall value appreciation. Moreover, new business opportunities arising from an ISMS-certified enterprise will contribute to enhancing transaction stability, and enhanced efficiency of operations within an ISMS-certified enterprise will result in an increase in trust. Finally, the protective capacity of an ISMS-certified enterprise will lead to a heightened level of security awareness in all their operations.
\section{Case Study}
\label{Researchmethod}

\subsection{Context}
\label{ResearchContext}
Kempower designs and manufactures fast-charging solutions for electric vehicles and machinery that utilize Direct Current (DC). Based in Lahti, Finland, the company sources over 90 percent of its materials and components locally \cite{kemppigroup}. Kempower's product range spans various transportation modes, including passenger cars, public transport, boats, and mining equipment. Kempower initiated an ISMS project to gain ISO 27001:2022 certification. Kempower’s ISMS project initiation consists of two phases. In the first phase, an as-is status assessment of Cyber Security Management was done against the ISO 27001 standard. In the second phase of the project, all employees participated in a Cyber Security Development kick-off  meeting, whose target was to increase Kempower employees' awareness of Cyber Security. The project was initiated  to develop all tools, technologies,  practices, and policies to meet ISO 27001:2022 requirements. One of the motivations behind this research is to investigate whether the use of AI can enhance work processes during the ISMS project.

\begin{comment}
    \begin{figure}
    \centering
    \includegraphics[width=1\linewidth]{kemppigroup.png}
    \caption{Kemppi Group company structure}
    \label{fig:enter-label}
\end{figure}
\end{comment}

\subsection{Research Motivation and Objective}
\label{ResearchObjective}

\textbf{Motivation}: The motivation for this research is driven by Kempower's need to establish its own ISMS. Implementing an ISMS not only enhances trust among business partners, thereby boosting operational efficiency, but it also contributes to a safer working environment by significantly reducing the number of security breaches.\cite{jalil2003isms}. Furthermore, this study explores the potential of AI to improve workflow processes during the ISMS project, thereby enhancing both the effectiveness and efficiency of these processes.

\textbf{Objective}: This research aims to evaluate the technical contributions of utilizing AI in the ISMS project at Kempower. Specifically, the study examines whether AI can play a beneficial role in this context. The investigation focuses on practical applications, such as the creation of ISMS documentation using innovative AI methods.

\subsection{Research Questions}
\label{ResearchQuestions}
We formulated three RQs in Table \ref{table:research-questions} based on the study objective. The responses to these RQs expected to enhance implementation AI-enhanced ISMS.

\begin{table}[t]
\centering
\caption{Research Questions and Rationale}
\label{table:research-questions}
\begin{tabular}{|>{\raggedright\arraybackslash}m{5cm}|>{\raggedright\arraybackslash}m{6.5cm}|}
\hline
\rowcolor{lightgray} \textbf{Research Questions} & \textbf{Rationale} \\ \hline
\textbf{RQ1}: %How can the information security management system be enhanced by leveraging AI in an organization 
How can an information security management system be enhanced by leveraging AI in an organization?
& This research questions aims to help with understanding how generative AI can possibly enhance the creation of an ISMS. The answer to this research question gives insight into how the use of AI can save resources and speed up the project for example. \\ \hline
\textbf{RQ2}: How do employees involved in cybersecurity projects perceive the concerns and disadvantages of leveraging AI? & This research questions aims to answer how different employees with differing backgrounds in using AI see the significance of using AI to help with this project. \\ \hline
\textbf{RQ3}: %Is it likely that K\textit{empower} could leverage AI in future projects? 
What are the broader implications of Kempower's use of AI for similar projects in the industry, and how can these insights benefit the scientific community?
& This research questions aims to help with getting an understanding if K\textit{empower} should use generative AI in the future while working on large scale projects like the ISMS project. \\ \hline
\end{tabular}

\end{table}

\subsection{Data Collection}
\label{RMandDC}
To address our research questions, we conducted semi-structured interviews. This format is particularly suitable for participants who are not professionals in the field of information technology, as it allows for follow-up questions and enables deeper exploration of the topic based on participants’ responses. Moreover, semi-structured interviews are often underutilized despite their potential to elicit more insightful and meaningful contributions from study participants, directly addressing the specific dimensions of the research questions \cite{galletta}.

\subsection{Data Collection Process}
\label{Interviews}
.
The interviews were conducted by the first author of the paper. The average length of the interviews was 12 minutes. The interviews were held in Finnish, as all the participants' first language is Finnish. The answers were later translated into English. Before the interviews started, the participants gave their permission to have their answers used in this study. The interviews were semi-structured, and they all were asked the same ten questions. In addition to the ten questions, there was room for discussing the topic further and exploring additional questions. The questions used in this particular interview were planned with the cyber security team, so that Kempower would benefit the most from this paper.The interview questions used are listed in Table \ref{tab:interview_questions}: These IQs are driven by the study's RQs. We have also mapped each RQ to the corresponding interview questions in Table \ref{tab:interview_questions}. For example, IQ2-IQ4 correspond to RQ1, focusing on how AI is used and its role in ISMS development. IQ5-IQ7 align with RQ2, exploring employee perceptions of AI, including benefits, disadvantages, and challenges. Lastly, IQ8-IQ10 relate to RQ3, examining the broader implications of AI in large-scale projects and its future potential.

\begin{table}[h]
\centering
\caption{Interview Questions for Information Security and AI}
\label{tab:interview_questions}
\begin{tabularx}{\textwidth}{|p{1cm}|X|p{1cm}|}
\hline
\rowcolor{lightgray} \textbf{\#} & \textbf{Interview Question} & \textbf{RQ} \\ \hline

IQ1 & Briefly introduce your experience in information security and AI. & - \\ \hline

IQ2 & How do you see AI being used in this project? & RQ1 \\ \cline{1-2}
IQ3 & How can you help with the development and maintenance of an ISMS? & \\ \cline{1-2}

IQ4 & What safeguards need to be in place when using AI tools in the maintenance of the ISMS?& \\ \cline{1-2}
IQ5 & What were the benefits of leveraging AI in the ISMS project?& \\ \hline

%IQ5 & What were the benefits of leveraging AI in the ISMS project? & RQ2 \\ \cline{1-2}
IQ6 & What about disadvantages? & RQ2\\ \cline{1-2}
IQ7 & What are the primary challenges and/or concerns when using AI in projects like this? & \\ \hline

IQ8 & Can you see Kempower using AI in other large-scale projects as a tool in the future? & RQ3 \\ \cline{1-2}
IQ9 & Did you notice that the documents were written with the help of AI? How? & \\ \cline{1-2}
IQ10 & Did you see AI making mistakes in the documents? & \\ \hline

\end{tabularx}
\end{table}

\subsection{Participant Selection}

\begin{figure}[t]
    \centering
    \includegraphics[width=1.0\linewidth]{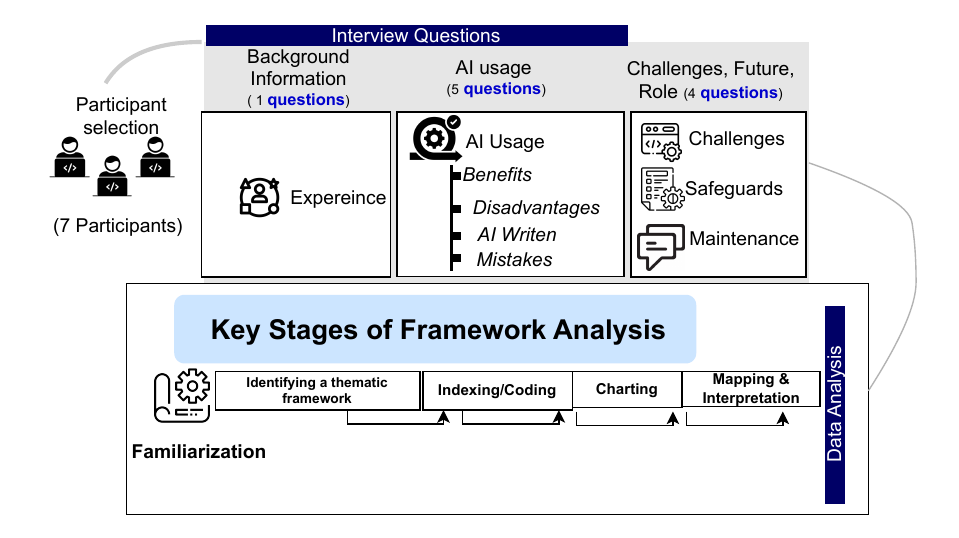}
    \caption{Overview of Survey Questions and Framework Analysis Method}
    \label{fig:stages}
\end{figure}

The participants are individuals designated by Kempower's management team as cyber-resource personnel from each business function at Kempower. They are responsible for reviewing all 116 different ISO 27001:2022 documents which makes them the best candidates for the interviews, as they are the ones who read and possibly provide comments on the documents. AI was heavily used during the development of these documents. The interviews seek to explore, among other aspects, how the participants perceive the significance of AI in the ISMS project and whether they believe AI could potentially be beneficial in other large-scale projects at Kempower in the future. The participants' experience in AI is listed in Table \ref{table:demographics}.

\begin{table}[t]
\centering
\caption{AI experience of the interviewed persons}
\label{table:demographics}
\begin{tabularx}{\textwidth}{|X|c|c|c|c|c|c|c|}
\hline
\rowcolor{lightgray} \textbf{Participants (P)} & \textbf{P1} & \textbf{P2} & \textbf{P3} & \textbf{P4} & \textbf{P5} & \textbf{P6} & \textbf{P7} \\ \hline
Experience in AI & x & & & & x & x & x \\ \hline
Experience in cyber security & x & & x & x & x & x & x \\ \hline
No experience & & x & & & & & \\ \hline
\end{tabularx}
\end{table}

\subsection{Data Analysis}
\label{DataAnalysis}

In this study, we employed framework analysis, a qualitative data analysis technique, to examine the data gathered from interviews. This method is particularly suitable for research that seeks to derive new insights from collected data, as opposed to testing established hypotheses or theories \cite{hassan}.
The key stages of framework analysis are depicted in Figure \ref{fig:stages}. These stages include familiarization, identifying a thematic framework, indexing, charting, and mapping and interpretation \cite{lacey2007qualitative}. When used effectively, the framework analysis method is a systematic and flexible approach to analyzing qualitative data \cite{gale2013using}.

\begin{itemize}
    \item \textbf{Familiarization Stage}: This initial stage involves either partial or full transcription of the data, followed by thorough reading.
    \item \textbf{Identifying a Thematic Framework}: The next step is to develop an initial coding framework. This framework is based on pre-existing issues and those that emerge during the familiarization stage of the analysis.
    \item \textbf{Indexing}: Once the thematic framework is identified, the indexing stage begins. This involves applying the thematic framework to the data using numerical or textual codes to identify pieces of data that correspond to specific themes. In this paper, the data were coded with colors rather than numbers or text to enhance the visualization of the data collected during the interviews.
    \item \textbf{Charting}: Several charts are created in this stage to facilitate analysis across the entire dataset. For each interview question, a chart was created, and every different answer theme was added as a column. This organization allowed for an easy comparison of how many participants shared the same answers, enhancing the readability of the data. 
    \item \textbf{Mapping and Interpretation}: The final stage involves searching for patterns, concepts, associations, and explanations within the data. The charts created in the previous stage aid in this process, serving as visual tools to display insights derived from the data.
\end{itemize}

\section{Results}
\label{Results}

\subsection{Enhancing the ISMS by Leveraging AI (RQ1)}
AI was utilized in creating the documentation, including the creation of control document templates and baselines. Because AI was used, the documentation did not start from scratch as AI made the first baseline version of the document. AI also helped with the structure of the documents (see Table 4). After these steps, Kempower specific information was added to the documents. This way the problem of a blank page was eliminated. Participant P4 did not have a clear understanding of how AI was used in this project. Participant P4 felt that AI was used to search for information and make manual labor faster. Participant P7 mentioned that: \faHandORight{} “\textit{In this project, AI was used innovatively, securely and efficiently}.”

\begin{table}[h]
\centering
\caption{Task distribution among participants}
\label{tab:task_distribution}
\begin{tabularx}{\textwidth}{|X|c|c|c|c|c|c|c|}
\hline
\rowcolor{lightgray} \textbf{Participant (P)} & \textbf{P1} & \textbf{P2} & \textbf{P3} & \textbf{P4} & \textbf{P5} & \textbf{P6} & \textbf{P7} \\ \hline
Generating the baseline for the documents & x & x & x &  & x &  & x \\ \hline
Creating the documentation &  & x &  & x & x &  &  \\ \hline
Searching for information &  & x & x &  & x &  & x \\ \hline
\end{tabularx}
\end{table}

The usage of AI in this project helped with time and cost savings. As mentioned in Table 5, with the help of AI, the documentation could be completed very quickly. There were also cost savings because there was no need to purchase help from external cybersecurity consultants, for example. AI helped with producing a lot of material in a short period of time. One other benefit was that the documentation did not need to start from scratch. AI can produce new insights into things that the employees working on the project did not think about. Participant P6 mentioned that the workload of employees has been lighter because AI was used, thus saving time for the people working with the documentation. Participant P6 mentioned that: 
\faHandORight{} “\textit{in the time it would take to complete one project without the assistance of AI, we can complete two projects with the help of AI}.”
\begin{table}[h]
\centering
\caption{Perceived benefits reported by participants}
\label{tab:perceived_benefits}
\begin{tabularx}{\textwidth}{|X|c|c|c|c|c|c|c|}
\hline
\rowcolor{lightgray} \textbf{Participant (P)} & \textbf{P1} & \textbf{P2} & \textbf{P3} & \textbf{P4} & \textbf{P5} & \textbf{P6} & \textbf{P7} \\ \hline
Time savings & x & x & x & x & x & x & x \\ \hline
Cost savings & x & x &  & x &  & x &  \\ \hline
Not having to start the documentation from scratch & x & x &  &  &  &  &  \\ \hline
New insights &  &  &  &  & x &  &  \\ \hline
\end{tabularx}
\end{table}
The participants mentioned that the material created with AI needs to be reviewed carefully by a human. Even if AI generates information that appears to be correct, a professional must always verify the output to reduce the risk of misinformation inside the organization. Employees must not share Kempower specific information with AI and they also should not copy anything from AI without verifying the information first. Participant P6 was concerned that the documentation will not be kept up-to-date and Kempower specific if it is maintained with the help of AI.

\begin{tcolorbox}[colback=lightgray, colframe=black!50, title=Takeaway 1, boxrule=0.5mm, coltitle=black, fonttitle=\bfseries, sharp corners=south]
Using AI in developing the ISMS has been shown to significantly enhance efficiency, as evidenced by the rapid production of initial document baselines and structures—this was echoed by Participant P6 who noted, “\textit{in the time it would take to complete one project without the assistance of AI, we can complete two projects with the help of AI}.” However, our findings also indicated the crucial role of human oversight in verifying AI-generated content for accuracy and relevancy. Participants reported that while AI accelerated the documentation process, every piece of AI-generated content required careful review to mitigate risks associated with automation.
\end{tcolorbox}

\subsection{The concerns and disadvantages of utilizing AI (RQ2)}
The participants thought that the primary challenges and concerns when using AI were that the material produced with AI needed to be reviewed very carefully. The control document materials need to align with Kempower's ways of working and the standard controls as closely as possible. Another concern with the use of AI was that employees can’t share insider information with AI as there is a risk of the information being leaked. Participant P4 was concerned that Kempower should not blindly follow the instructions AI gave because it does not have Kempower specific information.
\begin{table}[h]
\centering
\caption{Participants concerns regarding AI and information handling}
\label{tab:ai_concerns}
\begin{tabularx}{\textwidth}{|X|c|c|c|c|c|c|c|}
\hline
\rowcolor{lightgray} \textbf{Participant (P)} & \textbf{P1} & \textbf{P2} & \textbf{P3} & \textbf{P4} & \textbf{P5} & \textbf{P6} & \textbf{P7} \\ \hline
Validation of the material created with AI & x & x & x & x & x &  & x \\ \hline
Leaking of information, not sharing insider information with AI & x & x & x & x & x & x &  \\ \hline
AI can create false information &  &  &  & x &  &  &  \\ \hline
ChatGPT's knowledge is limited to events before 2021 &  &  &  &  &  &  & x \\ \hline
\end{tabularx}
\end{table}

Participant P4 mentioned that there should always be a human with experience in information security to verify and review the document which AI has made. Participant P5 was concerned that ChatGPT is not a closed environment and worried that the AI might learn confidential information if it is input as prompts. Participant P7 mentioned that ChatGPT provided too much information and that the prompts used needed to be anonymized. Participant P4 mentioned that:  \faHandORight{} \textit{“producing relevant information with the help of AI is a skill that must be learned"}.

\begin{table}[h]
\centering
\caption{Disadvantages of using AI according to the participants}
\label{tab:ai_criticisms}
\begin{tabularx}{\textwidth}{|X|c|c|c|c|c|c|c|}
\hline
\rowcolor{lightgray} \textbf{Participant (P)} & \textbf{P1} & \textbf{P2} & \textbf{P3} & \textbf{P4} & \textbf{P5} & \textbf{P6} & \textbf{P7} \\ \hline
The nature of work is more just reviewing the documents & x &  & x &  &  &  &  \\ \hline
Writing extra information which isn't Kempower specific & x &  &  & x & x &  &  \\ \hline
AI can produce misinformation &  & x & x &  &  &  &  \\ \hline
Risk of blindly following AI's suggestions &  &  &  & x &  &  &  \\ \hline
Not learning about the controls as much as without the help of AI &  &  &  &  &  &  & x \\ \hline
\end{tabularx}
\end{table}
The disadvantage of using AI in this project was that the nature of work shifted towards only reviewing the documentation created by AI rather than creating it ourselves. AI also generated a lot of information that was not needed in the documents and was not Kempower-specific. One disadvantage was that AI can produce false information. According to Participant P4, there is a risk that Kempower may blindly follow AI’s instructions rather than doing what is best for the organization. The concern of trying to find the relevant information which can be applied to the organization from the large amount of text produced by AI was brought up by participant P5. One concern that came up in the interviews was that the documentation was too easy to create with the help of AI and the employees working on the documentation did not learn so much about the content of the control documents (see Table 7).

%\begin{tcolorbox}[colback=lightgray, colframe=black!50, title=Takeaway 2, boxrule=0.5mm, coltitle=black, fonttitle=\bfseries, sharp corners=south]
%While AI significantly enhances efficiency in creating ISMS documentation, human oversight is essential to verify AI-generated content for accuracy and relevancy. Key concerns include the risk of information leakage, potential for AI to produce false or non-specific information, and the necessity for employees to understand and critically evaluate AI outputs to ensure alignment with organizational standards.
%\end{tcolorbox}

\begin{tcolorbox}[colback=lightgray, colframe=black!50, title=Takeaway 2, boxrule=0.5mm, coltitle=black, fonttitle=\bfseries, sharp corners=south]
While AI significantly enhances efficiency in creating ISMS documentation, its use introduces unique challenges that necessitate vigilant human oversight. Key concerns include the potential for information leakage, the risk of AI producing non-specific or inaccurate information, and the crucial need for employees to critically evaluate AI outputs to ensure they align with organizational standards and real-world applicability. Participants expressed that effectively utilizing AI requires a specialized skill set, including the ability to discern relevant from irrelevant AI-generated content and ensuring that internal knowledge is not compromised.
\end{tcolorbox}

\subsection{AI as a Tool in the Future (RQ3)}

All of the participants thought that Kempower can use AI in other projects as a tool in the future. One of the ways Kempower can leverage AI is that it can help with reducing the time it normally takes to gather information from the internet. Certain time-consuming tasks, such as data collection and measuring current status can be automated or made quicker with AI. Participant P5 mentioned that AI can be very helpful in projects where something concrete has already been established. AI may have difficulties in creating something such as different processes from scratch as they might not fit the organization’s needs. Participant P6 thought that AI should be used more at Kempower, as it was only being used at a basic level at the time of writing this paper. According to participant P7, AI is a tool that no other tool compares to when it comes to creating basic content for projects, such as documents. They think that using AI is an excellent way of working.

\begin{tcolorbox}[colback=lightgray, colframe=black!50, title=Takeaway 3, boxrule=0.5mm, coltitle=black, fonttitle=\bfseries, sharp corners=south]
Participants believe AI can significantly streamline future projects at Kempower by automating time-consuming tasks like data collection and basic content creation, though it may struggle with developing entirely new processes tailored to specific organizational needs.
\end{tcolorbox}
\section{Discussion}
\label{Discussion}

\subsection{Participants experience}
The participants in this study had varying levels of experience in cyber security. The participants' experience did not significantly impact their answers. Experience with AI did not vary as much as experience with cybersecurity. This may be due to the rise in popularity of AI in the recent years arguably due to the publication of ChatGPT.

\subsection{Utilization of AI}
\begin{figure}[t]
    \centering
    \includegraphics[width=1\linewidth]{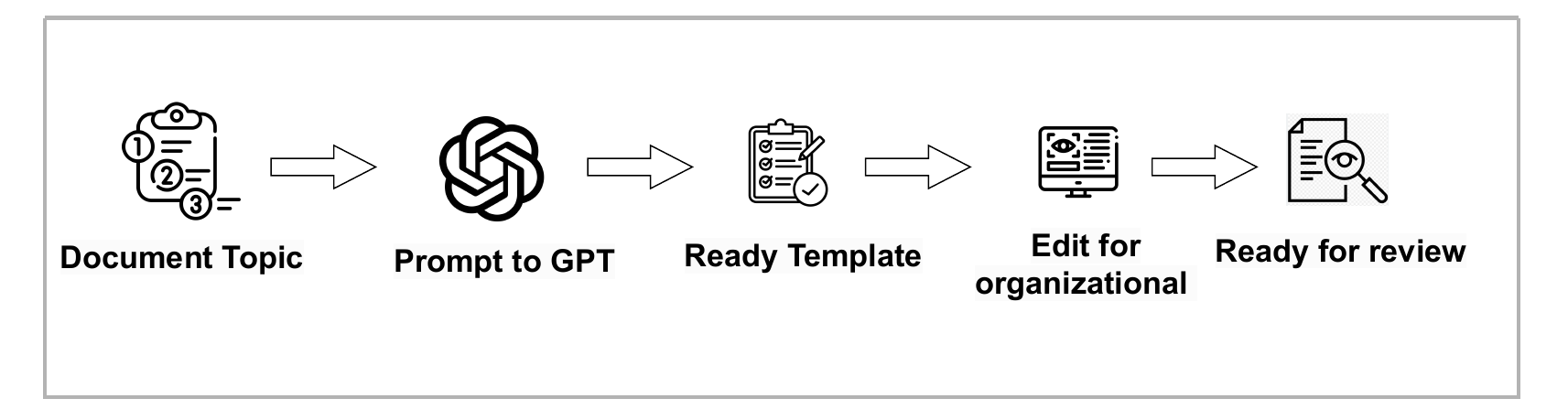}
    \caption{Utilization of AI in ISMS document creation}
    \label{fig:AI-and-ISMS}
\end{figure}

In this paper, there were clear answers to all of the research questions. The main research question was “How can the information security management system be enhanced by leveraging AI in an organization?" The participants who were a part of this study had multiple answers to this question such as that AI could be used in creating the documentation for the ISMS. AI was used in different phases of a document’s life cycle. AI was helpful in streamlining the process of searching for information by eliminating it altogether which made the documentation phase of the project a lot easier and faster as ISO27001:2022 contains 116 different controls which would otherwise require a lot of searching for information. With the help of AI, the first baseline or draft versions of the documents could be made within a couple of minutes. The creation process of ISMS documents when utilizing AI can be seen in Figure \ref{fig:AI-and-ISMS}.

\begin{figure}[t]
    \centering
    \includegraphics[width=1\linewidth]{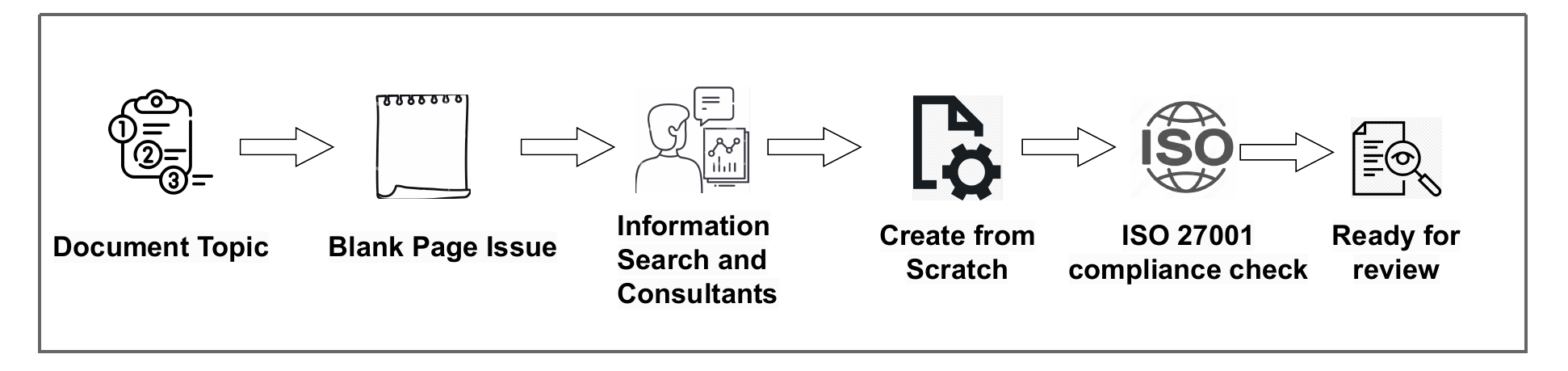}
    \caption{ISMS document creation without AI}
    \label{fig:ISMS-no-AI-label}
\end{figure}

As mentioned in the results of this paper, only the verification of the content created by AI needed to be done and modified to fit the ways of working of Kempower. Also, AI was helpful in eliminating the problem of a blank page. The participants also thought that AI can make certain time-consuming tasks quicker. The process of creating an ISMS document without the utilization of AI can be seen in Figure \ref{fig:ISMS-no-AI-label}.

One of the secondary research questions was “How the projects cyber resource employees perceive the significance of utilizing AI in a large scale project?” The projects cyber resource employees thought that there were far more benefits from using AI compared to the disadvantages. Interestingly, only a couple of the participants mentioned costs savings but all of them thought that the use of AI saved a significant amount of time. Cost savings were also achieved because, with the help of AI, there was no need to use external consultants. One could argue that time and cost savings go hand in hand as saving time also means saving resources in other areas as well. 

The disadvantages of using AI in this project, as listed by the participants, revolved around the fact that the nature of work shifted toward reviewing the material created by AI. The employees who were responsible for creating the documentation for the ISMS did not learn about the ISO27001 standard as much as they would have without the help of AI because using AI made writing the content of the documents so much easier. The participants also thought that AI can produce misinformation and it created a lot of information which was not Kempower specific. This was clearly an issue which the employees had to deal with because there is no way AI tools such as ChatGPT could know how Kempower’s different processes work. There was a concern that someone might share confidential information about Kempower to an AI tool.

Some of the participants noticed that AI was used in the creation of the documents because AI could not produce Kempower specific information. One could assume that AI did a very good job in creating the documents, as participants with experience in AI and cybersecurity did not notice its use. Some of the participants also stated that they did not actively think that the document they were reading were written with the help of AI, but there were some things that made it seem like it. AI does not make grammatical errors clearly as much as a human would make. More than half of the participants thought that AI made mistakes in the documents. The mistakes AI made was not because it did not know the ISO27001 standard but because the text it provided was superficial. One of the participants noticed structural errors in the documents which also does not mean that the AI lacks knowledge of the ISO27001 standard. These so-called mistakes reinforce the fact that there always has to be an experienced employee reviewing and verifying the material created with the help of AI.

The last research question in this paper was “Is it likely that Kempower could leverage AI in future projects?” There was a clear answer to this question, as all the employees who participated thought that Kempower could use AI in other large-scale projects in the future. In the interviews it was mentioned that Kempower should use AI tools more as it is only used at a really basic level at the moment. AI is an excellent tool that can be utilized in many ways across various projects. AI tools streamline different processes and enhance the overall efficiency of working. In the future, Kempower will only use AI tools that have a limited environment within the organization. One assumption that can be made from the interviews is that if Kempower had an AI tool in use which would be a closed environment, it would be even better as the employees would not have to worry about leaking insider information accidentally with AI.

The employees interviewed and their business functions are a valuable asset in the maintenance of the ISMS. 
Some of the participants are responsible for maintaining their own business functions documents in the ISMS. There were a couple of things that came up in the interviews about the safeguards that need to be in place when using AI tools in the maintenance phase of the ISMS. The concerns Kempower need safeguards against revolve around the fact that employees have to use AI safely and not share insider information to AI tools as well as there has to be an experienced employee reviewing the material even in the maintenance phase.

\subsection{Reliability, Generalizability and Limitations of the Study}
Some parts of this paper can be generalized. There is no doubt that AI can be utilized in the creation of an ISMS in other organizations as well. As can be seen from the results of this paper, Kempower should use AI in other projects similar to the ISMS project as well. One could make a conclusion from this point that other organizations could use AI in their projects as well, be it an ISMS project or other similar project. There is problems with generalizing this paper to a wider audience because the scope of this paper was only Kempower. There is no guarantee that other organizations are implying AI to their day to day working life and this paper is of no value for such organizations if they are not seeking to change their ways of working. One point to draw out of the reliability of this paper is that when this paper and the ISMS project was started, ChatGPT’s knowledge was limited to events that happened before the year 2021. This can be seen as a problem as the version of ISO27001 standard Kempower made their ISMS according to was published in 2022. This does not have a large impact on the quality of the ISMS because as can be seen from the results of this paper, an experienced employee always had to validate the material which was created with the help of AI.

One limitation of this paper is that it only studied a specific organization's ISMS project. The results could be generalized to a wider audience if there were other organizations with their own ISMS projects studied. This would lead to a wider sample size which would create a more reliable study and thus more reliable results. One could argue that a limitation of this paper is that the participants chosen for this study may perceive the usefulness of AI differently than others. This would not be an issue if the sample size would have been bigger. Only some of the participants chosen to this study produced documents or other material in to the ISMS. Some of the participants only went trough and read the material created with the help of AI which may have an impact on the papers results as they did not use AI themselves during the ISMS project.
\section{Conclusion}
\label{Conclusion}
The objective of this paper was to study how AI usage can help in the creation process of Kempower’s ISMS. The research questions were answered by conducting multiple interviews. The participants had differing backgrounds and knowledge of information security and AI which made the answers of the interviews more diverse. The most important result of this study was that the use of AI in the process of creating an ISMS from scratch not only makes the process easier but it also makes it quicker and saves resources. The resource savings were both internal and external, as Kempower employees had more time to work on other projects and there was no need for external consultant help in the documentation creation phase. In addition, there were many more benefits to using AI in Kempower’s ISMS project compared to the disadvantages. 
One major result of this paper was that AI usage made the ISMS project possible in such a short amount of time. The most notable disadvantage was that employees using AI had to be careful when inputting information into it. This is no longer an issue as Kempower will use closed environment GenAI/LLM models which are limited to information inside the organization moving forward. Because AI was used in the project, the work focused more on verifying material created by AI rather than employees creating it from scratch. This leads to the fact that employees might not learn as much about the ISO 27001:2022 standard when compared to the process of employees having to study and search for information themselves.

\bibliographystyle{splncs04}
\bibliography{refs}

\end{document}